\documentstyle[aps,prd,epsf,floats]{revtex}

\def\plb#1{Phys.\ Lett.\ {\bf B\,#1}}
\def\prd#1{Phys.\ Rev.\ {\bf D\,#1}}

\def\epjc#1{Eur.~Phys.~J.\ {\bf C\,#1}}

\newcommand{\beq}{\begin{equation}}
\newcommand{\eeq}{\end{equation}}
\newcommand{\beqs}{\begin{eqnarray}}
\newcommand{\eeqs}{\end{eqnarray}}

\newcommand{\BR}{\mbox{{\rm BR}}}

\newcommand{\MuToEee}{\mu^- \to e^- \, e^+ \, e^-}
\newcommand{\TauToEee}{\tau^- \to e^- \,  e^+ \,  e^-}

\newcommand{\andthis}{~~~~~~~\mbox{and}~~~~~~~} 

\draft

\begin{document}
\twocolumn[\hsize\textwidth\columnwidth\hsize\csname
@twocolumnfalse\endcsname

\draft
\vglue 0.8cm
\title{On the massless ``just-so" solution     
 to the solar neutrino problem    
\vglue -2.2cm
{\small \hfill {hep-ph/0012089}\\
        \hfill  { } 
}
\vglue 1.0cm
}
\author{M.\ M.\ Guzzo$^1$, H.\ Nunokawa$^1$, P.\ C.\ de Holanda$^1$ and 
O.\ L.\ G.\ Peres$^{1,2}$}
\address{$^1$Instituto de F\' {\i}sica Gleb Wataghin, 
    Universidade Estadual de Campinas, UNICAMP\\    
    13083-970 Campinas SP, Brazil\\
$^2$The Abdus Salam International Center for Theoretical Physics, I-34100
Trieste, Italy
}

\maketitle
             
\begin{abstract}
We study the effect of the non-resonant, vacuum oscillation-like
neutrino flavor conversion induced by non-standard flavor 
changing and non-universal flavor diagonal neutrino interactions 
with electrons in the sun. 
We have found an acceptable fit for the combined analysis for the 
solar experiments total rates, the Super-Kamiokande (SK) energy spectrum 
and zenith angle dependence. Phenomenological
constraints on  non-standard flavor 
changing and non-universal flavor diagonal neutrino interactions are
considered. 
\end{abstract}

 
\vskip2.0pc]

%

%
Non-standard neutrino interactions with matter can generate neutrino
flavor oscillations. This phenomenon was suggested by Wolfenstein in
his seminal 1978 paper \cite{W}.  
Applications of this idea to the solar neutrino problem were first 
suggested in 1991~\cite{GMP,ER} when it was observed that resonantly 
enhanced neutrino oscillations induced by non-standard neutrino 
flavor changing (FC) as well as non-universal flavor diagonal (FD) 
neutrino interactions can explain the solar neutrino experimental 
data~\cite{solarnuexp}  which clearly indicates a solar neutrino flux
smaller than  what is predicted by the standard solar 
models~\cite{BP98}.

Interestingly enough, such oscillations can be resonantly enhanced 
even if neutrinos are massless  and no vacuum mixing angle 
exists~\cite{GMP}, as a result of an interplay between 
the standard  electroweak neutrino charged currents 
and non-universal non-standard flavor diagonal neutrino 
interactions with matter. 
In fact, in this mechanism, resonance plays a crucial role 
in order to provide a viable solution to the solar neutrino 
problem \cite{BPW,KB97,BGHNK}.

It should be emphasized that if such nonstandard neutrino 
FC and FD interactions exist only with electrons, 
no resonant conversion can happen because 
the mixing angle in matter is constant, as we will see later, 
contrary to the case of usual MSW effect~\cite{msw}, or the
case with $d$,$u$-quarks FC and FD interactions. 	
From this point of view, the oscillation induced by 
non-standard neutrino interactions with electrons alone 
is similar to the vacuum oscillation mechanism
despite the difference that it occurs only in matter, 
inside the sun.

This non-resonant neutrino conversion can be useful to explain the
solar observations. The first discussion on this possibility appeared
in Ref. \cite{BPW}  where non-resonant neutrino oscillation induced
by FC and FD interactions only with electron in the solar matter was
mentioned as a possible solution to the solar neutrino problem. 

Nevertheless, so far, no quantitative analysis of this 
scenario was presented.

In this brief report we investigate this 
possibility by performing a detailed fit to 
the most recent solar neutrino data. 
We conclude that non-resonant neutrino oscillations induced 
by non-standard neutrino interactions can only 
provide a rather poor fit to the total rates observed by 
all the solar neutrino experiments coming from 
Homestake, Gallex/GNO, Sage and SK~\cite{solarnuexp}
whereas when we include also the full SK 
recoil electron spectrum and the zenith angle dependence
the fit become an acceptable one. 
We find also that this fit requires the 
new non-standard neutrino interactions parameters to be very large.

Here we assume that neutrinos have non-standard FC as well as FD 
interactions only with electrons which could be realized 
in some models such as minimal super-symmetric standard 
model without R-parity \cite{r-parity} or 
$SU(3)_C \otimes SU(3)_L \otimes U(1)_N$ (331) models \cite{331}. 
Under this assumption, the evolution equation for massless neutrinos 
in matter can be expressed as \cite{GMP}, 

\beq \label{eq-of-motion} 
i\frac{d}{dr} \pmatrix{A_e(r) \cr A_\ell(r)} = 
\sqrt2 G_F n_e(r)
\pmatrix{1                       & \epsilon_{el}  \cr
         \epsilon_{el} & {\epsilon'_{el}} \cr} 
\, \pmatrix{A_e(r) \cr A_\ell(r)} \,,
\eeq
where $A_e(r)$ and $A_\ell(r)$ ($l= \mu, \tau$) 
are, respectively, the probability
amplitudes to detect a $\nu_e$ and $\nu_\ell$ at position $r$ and
\beq \label{defeps}
\epsilon_{el} \equiv 
 {G_{\nu_e \nu_\ell} \over G_F} \andthis
{\epsilon}_{el}' \equiv 
 {G_{\nu_\ell \nu_\ell} - G_{\nu_e \nu_e} \over G_F} \,,
\eeq
describe, respectively, the relative strength of the FC and the FD
(but non-universal) interactions where $G_{\nu_\alpha \nu_\beta}$
($\alpha, \beta = e, \mu, \tau$)  denotes the effective coupling of the
respective interaction.

In this mechanism the mixing angle in matter $\theta_m$ does 
not depend on the electron density and is simply given by, 
%
\begin{figure}[t]
\epsfxsize=240pt
\hbox to\hsize{\hss\epsfbox{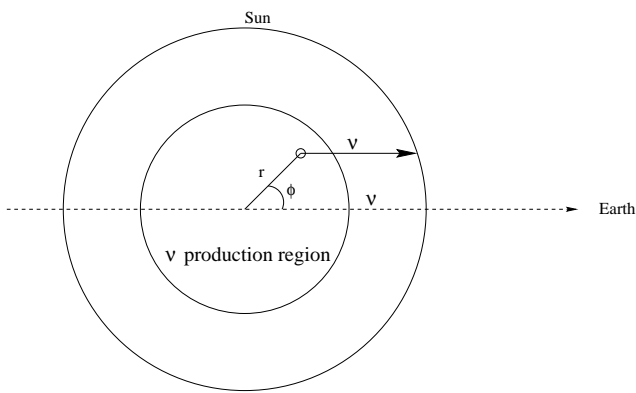}\hss}
\smallskip
\vglue 0.2cm 
\caption{
Definitions of the variables $r$ and 
$\phi$. (Size of the neutrino production region 
was enlarged just for the purpose of illustration.) 
} 
\label{fig:production}
\vglue -0.2cm 
\end{figure}

\begin{equation}
\sin^2 2\theta_m = 
\frac{4 \epsilon^2}{(1-\epsilon')^2 + 4 \epsilon^2}.
\label{eq:angle}
\end{equation}
We see that no MSW like resonance can occur because the
mixing angle in matter is constant and does not change 
along the neutrino trajectory 
(however see Ref.~\cite{botella}).

Let us introduce the two variables $r$ and $\phi$ which 
define the production point of neutrinos in the sun
as shown in Fig.~\ref{fig:production}. 
Then, for given values of ($\epsilon,\epsilon'$) and a given
production point in the sun  defined by $r$ and $\phi$
the survival probability of  electron neutrinos at the 
solar surface can be written as \cite{BPW}, 
\begin{equation}
\label{prob}
P(\nu_e \to \nu_e;r,\phi) = 1- \sin^2 2\theta_m 
\sin^2 \frac{\Psi(r,\phi)}{2},
\end{equation}
where 
\begin{equation}
\Psi(r,\phi) \equiv \sqrt{4 \epsilon^2 + (1-\epsilon')^2}
\sqrt{2}G_F\int_0^{x_{max}}  
\hskip -0.2cm
N_e(r,\phi,x) dx,
\end{equation}
where the $N_e(r,\phi,x)$ is the electron density 
profiles along the neutrino trajectory which starts 
at the creation point $(r,\phi$) corresponding to 
$x=0$ and ends at the solar surface 
corresponding to $x=x_{max}$. 
Note that there is no energy dependence in the 
probability.

From Eq. ({\ref{prob}) we can estimate the 
oscillation length as 
\begin{eqnarray}
L_{osc} 
&& 
\equiv \frac{2 \pi}
{\sqrt{2}G_FN_e\sqrt{(1-\epsilon')^2 + 4\epsilon^2}} \nonumber \\
&& 
\simeq  \frac{2.4\times 10^2}
{\sqrt{(1-\epsilon')^2 + 4\epsilon^2}}\ 
\left[ \frac{65\ \mbox{mol/cc}}{N_e} \right]
\  \mbox{km},
\label{osclength}
\end{eqnarray}
where we take $N_e = N_e(R \simeq  0.1 R_\odot) \simeq 65$ mol/cc as 
a reference value. 
{}From Eq. (\ref{osclength}) we see that if either 
$|1-\epsilon'|$ or $|\epsilon |$ is of the order of   $0.01$,  
the oscillation length 
is typically less than a few percents of the solar radius 
in the neutrino production region.  
This implies that for such values of $\epsilon$ and $\epsilon'$ 
there are many oscillations before neutrinos reach the solar 
surface and the final survival probability which is averaged 
over the neutrino production point will be, 
\begin{equation}
\langle P(\nu_e \to \nu_e) \rangle 
\simeq  1- \frac{1}{2}\sin^2 2\theta_m, 
\end{equation}
for any values of $r$ and $\phi$ and therefore, 
for any sources of neutrinos~\cite{BPW}.  
Therefore, such rapid oscillation 
cannot fit the solar neutrino data well.

As pointed out in Ref. \cite{BPW}, an 
interesting possibility remains if both 
$|1-\epsilon'|$ and $|\epsilon |$ is smaller than $\sim 0.01$. 
For such $\epsilon'$ and $\epsilon$,  
if $\Psi \sim (2n+1)\pi$  with small $n$, neutrinos 
produced as $\nu_e$ can be almost $\nu_x$ ($x = \mu, \tau$)
at the solar surface. 
On the other hand, if $\Psi \sim 2n\pi$ 
$\nu_e$ remains as $\nu_e$ at the solar surface. 

Since neutrinos from different nuclear reaction origins 
have different production distributions,  
there is a possibility that properly adjusting the 
parameter $\epsilon$ and $\epsilon'$ 
neutrinos from different reaction origins 
could be oscillated into another flavor in such 
a way that the solar neutrino data can be accounted for. 
In principle, this could happen if neutrinos oscillate 
only once or a few times before they reach the solar surface, 
similar to what happen to 
the case of the vacuum long-wavelength 
oscillation solution to the solar neutrino 
problem \cite{just-so}. 

In order to settle this issue we have performed a detailed 
$\chi^2$ analysis using the latest standard solar model by Bahcall
{\it et al.}~\cite{BP98} (BP98 SSM) as well the latest results 
of the current solar neutrino experiments coming from 
Homestake, Gallex/GNO, SAGE and SK \cite{solarnuexp}. 

The fitting procedure is as follows. 
We first compute the survival probability of $\nu_e$ at the
solar surface by the formula in Eq.(\ref{prob}) 
for various different production points defined by 
$r$ and $\phi$ as in Fig.~\ref{fig:production}.  
Then we compute averaged survival probabilities for 
neutrinos from different sources, i.e., 
$pp$, $pep$, $^7$Be, $^8$B, $^{13}$N 
and  $^{15}$O (we neglect other minor contributions from 
$^{17}$F and $hep$ for simplicity)  
taking into account the neutrino production point distribution 
from BP98 SSM.
After we compute the expected solar neutrino signal for 
each experiment, using the survival probability obtained above,  
we perform a $\chi^2$ analysis 
following the prescription given in Ref. \cite{fogli}. 

\begin{figure}[t]
\vglue -0.7cm 
\epsfxsize=260pt
\hbox to\hsize{\hss\epsfbox{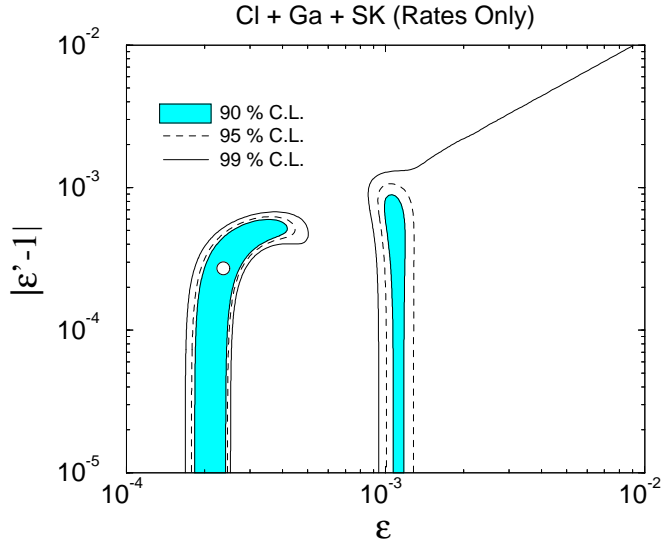}\hss}
\smallskip
\vglue -0.3cm 
\caption{
Allowed parameter region.
Region allowed by the total rates.
Best fit is obtained when $|\epsilon | = 2.4 \times 10^{-4}$ 
and $|\epsilon'-1| = 2.7 \times 10^{-4}$ with 
$\chi^2_{min} = 7.5 $ for 2 d.o.f. 
} 
\label{fig:allowed_fix}
\end{figure}

We show in Fig.~\ref{fig:allowed_fix}
the allowed parameter region determined 
by our $\chi^2$ analysis. 
We have used only the total observed rates 
of solar neutrinos by four experiments. 
The best fit is obtained at $(|\epsilon |, |\epsilon'-1| ) = 
(2.4, 2.9) \times 10^{-4}$ with 
$\chi^2_{min} = 7.5$ for 4 (data) -2 (free parameters) =  2 d.o.f. 
which corresponds to 2.4 \% C.L. indicating a poor fit.  
This is because the integrations of the survival probability 
over the variables $r$ and $\phi$ tend to kill the 
just-so suppressions of the neutrino fluxes and 
the final averaged probabilities from different sources 
end up with rather similar values to each other. 

However, the situation is still better than energy independent 
suppression since the following relation for the final 
averaged survival probability, 
\beq 
\label{P_relation}
\langle P(^8\text{B}) \rangle < 
\langle P(^7\text{Be}) \rangle < 
\langle P(pp) \rangle \,, 
\eeq
holds in this mechanism. 
In fact at the best fit point, we have obtained 
$\langle P(^8\text{B}) \rangle \sim 0.42$, 
$\langle P(^7\text{Be}) \rangle \sim 0.46 $ 
and $\langle P(pp) \rangle \sim 0.57$.

We also performed a $\chi^2$ analysis allowing $^8$B flux 
to vary freely but we do not find any significant 
improvement of the fit. 

Let us to include also the spectrum and zenith angle
dependence in our fit. 
First, let us note that this mechanism does not distort 
the SK energy spectrum since the conversion probability 
is completely energy independent. Therefore, 
contribution in $\chi^2$ from the spectrum is just constant
(does not depend on $\epsilon'$ and $\epsilon$) and 
the allowed parameter region shown in Fig. 2 is not affected
by the spectrum. 
Second, for the range of parameters we are considering in this work, 
there would be no significant effect of the earth matter
because the oscillation length in the earth is much larger 
than the earth radius. So again, 
contribution in $\chi^2$ from the zenith angle dependence is also 
just constant and the allowed parameter region shown in 
Fig. 2 is not affected by the zenith angle dependence. 

The total combined $\chi^2_{min}$ can be computed by simply 
adding the two constant contributions from spectrum and zenith angle
without affecting the allowed parameter region presented in Fig. 2. 
In this case, despite the fact that the fit is poor only with 
the total rates, 
when we combined SK spectrum as well as zenith angle dependence, 
we have obtained  $\chi^2_{min} = 25.6$ for 24 d.o.f. 
which corresponds to 34.4 \% C.L..

Here let us consider, as an interesting exercise, that the 
case when the systematic error of the Homestake is assumed 
to be 3 times larger than it has been reported. 
In Fig. 3 we present the region allowed by the rates 
under this assumption. We have obtained $\chi^2_{min} 
= 3.3$, for rates only which indicate the significant improved 
over the case presented in Fig. 2. 
We notice that this kind of exercise could be worthwhile to 
consider to take into account the possibility of some unknown
systematic effect of the Homestake experiment as it has not 
been calibrated with a radioactive source.

\begin{figure}[t]
\vglue -0.7cm 
\epsfxsize=260pt
\hbox to\hsize{\hss\epsfbox{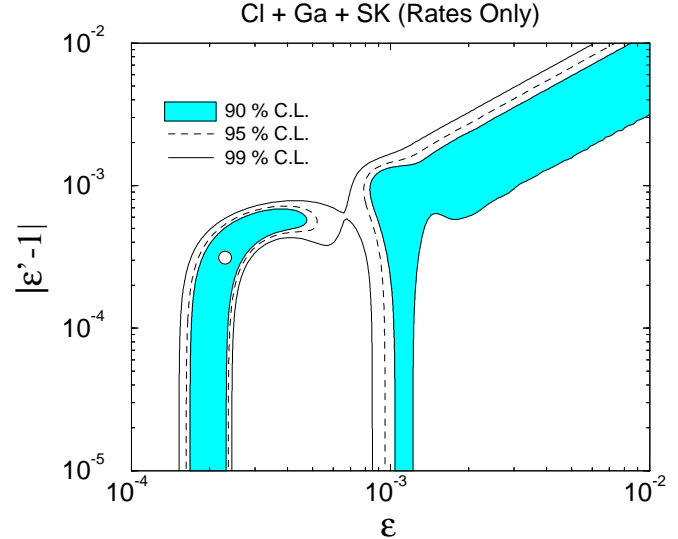}\hss}
\smallskip
\vglue -0.3cm 
\caption{
Same as in Fig. 2 but with the systematic error of the 
Homestake experiment was assumed to be 3 times larger.  
Best fit is obtained when $|\epsilon | = 2.3 \times 10^{-4}$ 
and $|\epsilon'-1| = 3.1 \times 10^{-4}$ with 
$\chi^2_{min} = 3.3 $ for 2 d.o.f. 
} 
\label{fig:allowed_fix_cl}
\end{figure}


The main problem appearing in the solution to the solar neutrino
problem based on  FC and non-universal FD neutrino interaction with
electrons is related with the magnitude of the  phenomenologically
required non standard parameters. 
Our  statistical analysis shows that although the FC parameter  
$\epsilon$ does not need to be very high,
($\epsilon_{el}\approx 10^{-3}$), the non-universal FD parameter
$\epsilon_{el}'$ is found to be of order of $1$.

The value for the FC
parameter $\epsilon$ is compatible with the available phenomenological
tests to the flavor conservation law. In fact, the most stringent
constraints on this parameter are due to the upper bounds on
$\MuToEee$ and $\TauToEee$~\cite{PDG}:
\beqs
\BR(\MuToEee)  &<& 1.0 \times 10^{-12} \,, \label{MuEbound} \nonumber \\
\BR(\TauToEee) &<& 2.9 \times 10^{-6}  \,, \label{TauEbound}
\eeqs
at  $90\%$ C.L.. %
Normalizing the above bounds to the measured rates of the related lepton
flavor conserving decays, $\BR(\mu^- \to e^- \, \bar\nu_e \,
\nu_\mu) \approx 100\,\%$ and $\BR(\tau^- \to e^- \, \bar\nu_e \,
\nu_\tau) = 0.1781$~\cite{PDG}, we obtain~\cite{BGHNK}
\beqs
\epsilon_{e\mu} \equiv G_{e \mu} / G_F  &<& 
 1.0 \times 10^{-6} \,, \label{GeBoundMu} \nonumber \\
\epsilon_{e\tau} \equiv G_{e \tau} / G_F &<& 
 4.2 \times 10^{-3} \,. \label{GeBoundTau}
\eeqs

Note, furthermore, that these bounds on $\epsilon$ 
can be also relaxed by a factor of 5-6 due to the breaking of the
$SU(2)_L$ symmetry~\cite{BGHNK}. Therefore, assuming that the neutrino
transitions involve the first and the third families, the required
value of $\epsilon$ is compatible with the phenomenological limits.

The challenge to  this solution is related with the required value of
the parameter $\epsilon_{el}'$ since  universality experimental tests
in the leptonic sector are very much stringent. 
In Ref.~\cite{BG_atm} the constraints involving the second and third
lepton families, i.e., interactions involving transitions of the type
$\nu_\mu\leftrightarrow \nu_\tau$ are obtained. 
It was found that~\cite{BG_atm}, 
\begin{equation}
\epsilon'_{\mu\tau} < 3.8\times 10^{-3}.
\label{eps_mu_tau}
\end{equation}

Note, however, that the parameter relevant for our present analysis of
the solution to the solar neutrino problem involves necessarily the
first neutrino family ($\nu_e$). Such constraint can be obtained
following the same steps of Ref.~\cite{BPW}. No direct limit can
be obtained to $\epsilon'_{e\tau}$. Nevertheless, since
$\epsilon'_{e\tau}=\epsilon'_{e\mu}-\epsilon'_{\mu\tau}$, limits on
this parameter are found considering the experimental constraints of
Eq.~(\ref{eps_mu_tau}) and limits on $\epsilon'_{e\mu}$. 

Non-zero values for $\epsilon_{ee}$ 
($\epsilon_{\mu\mu}$) gives a additional
contribution to $\nu_{e} e \rightarrow  \nu_{e} e$ ($\nu_{\mu} e
\rightarrow  \nu_{\mu} e$) cross section and can put constraints 
on $\epsilon_{ee}$  and $\epsilon_{\mu\mu}$~\cite{BPW}. We use the
more recent data about the $\nu_{e} e \rightarrow  \nu_{e} e$ total cross
section~\cite{lsnd2000}. 
This cross section is a function of $\epsilon_{el}$ and
$G_{\nu_e\nu_e}^A/G_F$ (the axial part of the effective coupling of the
respective interaction). 
We obtained 
\begin{eqnarray}
-2.56 <G_{\nu_e\nu_e}/G_F <0.63, 
\label{eq:1} 
\end{eqnarray}
at $90\%$ C.L. for arbitrary $G_{\nu_e\nu_e}^A/G_F$.
Taking the value quoted by Ref.~\cite{just-so}: 
$-0.18 <G_{\nu_{\mu}\nu_{\mu}}/G_F <0.14$. 
We obtained that $\epsilon_{e\mu}'$ is bounded to 
\begin{eqnarray}
-0.81 <\epsilon_{e\mu}' < 2.70,  
\label{eq:3} 
\end{eqnarray}
where the limit is at
$90\%$ C.L. .Using the constraint from Eq.~(\ref{eps_mu_tau})
and from Eq.~(\ref{eq:3}), we get finally,
\begin{eqnarray}
-1.81 <\epsilon_{e\tau}'-1 < 1.70,   
\end{eqnarray}
at $90\%$ C.L. .
From this constraint, we conclude that is possible to satisfy the
experimental constraints of FD couplings and at same time to be
compatible with the allowed region of the solar neutrino analysis. 
A additional caution is necessary because the same FD couplings that
induce neutrino oscillations can also change the detection cross
section, ($\sigma(\nu_{e} e \rightarrow  \nu_{e} e)$)  used for SK.
We check that the
absolute values of elastic cross section inside the range given in
Eq.(~\ref{eq:1}) are compatible with the assumed theoretical errors
used in the solar neutrino analysis. 
Also the shape of recoil electron in SK is not changed 
significantly due the FD couplings.

Concluding, we showed here for the first time a quantitative analysis of
non-standard flavor changing and non-universal flavor diagonal
neutrino interactions with electrons as a possible candidate 
to solve the solar neutrino
problem. If the parameters, $|{\epsilon}'_{el}-1| \ll 1$ and 
${\epsilon}_{el} \approx 10^{-4}-10^{-3}$ then we can get an acceptable 
fit for 
the combined analysis of the solar experiments total rates, the SK 
energy spectrum and the SK zenith angle dependence. We
conclude that the constraints on the violation of universality 
allow us a small ${\epsilon}_{e\tau}$ and a large value for 
${\epsilon}_{e\tau}'$ and is compatible with the preferred values
of our solar neutrino analysis. 
In this solution,  no spectrum distortion, no zenith angle dependence
and no seasonal effects are expected. Also only negative results are
expected in long-baseline experiments due the very large 
oscillation length.


\section*{Acknowledgments}

This work was supported by Funda\c{c}\~ao de Amparo \`a Pesquisa
do Estado de S\~ao Paulo (FAPESP), by Conselho Nacional de 
Ci\^encia e Tecnologia (CNPq) and 
by the European Union TMR network ERBFMRXCT960090.\vglue -0.4cm



\begin{references}
\bibitem{botella} %
If we taken in account the 1-loop corrections of the SM amplitudes, 
considered in 
%
F. J. Botella, C. S. Lim and W. J. Marciano,
\prd{35}, 896 (1987), 
%
then the mixing angle $\theta_m$ is not constant.
However this effect is very small and we do not
consider it here.
\end{references}
\end{document}